%% file: ARCHIVE.tex
\documentclass[sigconf]{acmart}

\usepackage{float}
\usepackage{balance}
\usepackage{amsmath}
\usepackage{mathtools}

\usepackage{url}
\usepackage{graphicx}     % advanced figures
\usepackage{xspace}     % fix space in macros
\usepackage{booktabs}     % nicer tables
\usepackage{enumerate}
\usepackage{amsmath, amssymb}
\usepackage[vlined,linesnumbered,ruled,noend]{algorithm2e}
\usepackage[caption = false]{subfig}
\usepackage{tabularx}
\usepackage{balance}
\usepackage{multirow}
\usepackage{amssymb}% http://ctan.org/pkg/amssymb
\usepackage{pifont}% http://ctan.org/pkg/pifont
\usepackage{adjustbox}
\usepackage{array}
\usepackage{xcolor,colortbl}
\usepackage{rotating}
\usepackage{siunitx}

\newcommand{\eg}{\textit{e.g.},\xspace}
\newcommand{\ie}{\textit{i.e.},\xspace}

\AtBeginDocument{%
  \providecommand\BibTeX{{%
    \normalfont B\kern-0.5em{\scshape i\kern-0.25em b}\kern-0.8em\TeX}}}

\renewcommand\footnotetextcopyrightpermission[1]{} % removes footnote with conference information in first column
\begin{document}

%%
%% The "title" command has an optional parameter,
%% allowing the author to define a "short title" to be used in page headers.
\title{I call BS: Fraud Detection in Crowdfunding Campaigns}

%%
%% The "author" command and its associated commands are used to define
%% the authors and their affiliations.
%% Of note is the shared affiliation of the first two authors, and the
%% "authornote" and "authornotemark" commands
%% used to denote shared contribution to the research.
\author{Beatrice Perez}
\email{beatrice.perez.14@ucl.ac.uk}
\affiliation{%
  \institution{University College London}
   \city{London}
  \country{UK}
}

\author{Sara R. Machado}
\email{s.machado@lse.ac.uk}
\affiliation{%
  \institution{London School of Economics}
    \city{London}
  \country{UK}
}

\author{Jerone T. A. Andrews}
\email{jerone.andrews@cs.ucl.ac.uk}
\affiliation{%
   \institution{University College London}
   \city{London}
   \country{UK}
}

\author{Nicolas Kourtellis}
\email{nicolas.kourtellis@telefonica.com}
\affiliation{%
  \institution{Telefonica Research}
  \city{Barcelona}
  \country{Spain}
}

%%
%% By default, the full list of authors will be used in the page
%% headers. Often, this list is too long, and will overlap
%% other information printed in the page headers. This command allows
%% the author to define a more concise list
%% of authors' names for this purpose.
\renewcommand{\shortauthors}{Perez and Machado, et al.}

%%
%% The abstract is a short summary of the work to be presented in the
%% article.
\input{sections/00_abstract}

%%
%% The code below is generated by the tool at http://dl.acm.org/ccs.cfm.
%% Please copy and paste the code instead of the example below.
%%


%%
%% Keywords. The author(s) should pick words that accurately describe
%% the work being presented. Separate the keywords with commas.
\keywords{Crowdfunding, Fraud, NLP, Image Processing, Deep Learning}

%%
%% This command processes the author and affiliation and title
%% information and builds the first part of the formatted document.
\maketitle

\input{sections/01_introduction}

\input{sections/02_relatedWork}

\input{sections/03_methodology}

\input{sections/04_texts}

\input{sections/05_images}
\input{sections/06_modeling}
\input{sections/08_conclusion}

\begin{acks}
This project has received funding from the European Union’s Horizon 2020 Research and Innovation program under the Marie Skłodowska-Curie ENCASE project (Grant Agreement No. 691025), CONCORDIA project (Grant Agreement No. 830927), and LIFECHAMPS project (Grant Agreement No. 875329). The paper reflects only the authors views and the Commission is not responsible for any use that may be made of the information it contains.
J. T. A. Andrews is supported by the Royal Academy of Engineering (RAEng) and the Office of the Chief Science Adviser for National Security under the UK Intelligence Community Postdoctoral Fellowship Programme.
\end{acks}

%% The next two lines define the bibliography style to be used, and
%% the bibliography file.
\bibliographystyle{ACM-Reference-Format}
\bibliography{sample-base}

\appendix
\section{Data Availability} 
The data that support the findings of this study are available from the corresponding author, B. Perez, upon reasonable request.

\end{document}

%% file: sections/00_abstract.tex
\begin{abstract}

Donations to charity-based crowdfunding environments have been on the rise in the last few years.
Unsurprisingly, deception and fraud in such platforms have also increased, but have not been thoroughly studied to understand what characteristics can expose such behavior and allow its automatic detection and blocking.
Indeed, crowdfunding platforms are the only ones typically performing oversight for the campaigns launched in each service.
However, they are not properly incentivized to combat fraud among users and the campaigns they launch:
on the one hand, a platform's revenue is directly proportional to the number of transactions performed (since the platform charges a fixed amount per donation);
on the other hand, if a platform is transparent with respect to how much fraud it has, it may discourage potential donors from participating.

In this paper, we take the first step in studying fraud in crowdfunding campaigns.
We analyze data collected from different crowdfunding platforms, and annotate 700 campaigns as \textit{fraud} or not.
We compute various textual and image-based features and study their distributions and how they associate with campaign fraud.
Using these attributes, we build machine learning classifiers, and show that it is possible to automatically classify such fraudulent behavior with up to 90.14\% accuracy and 96.01\% AUC, only using features available from the campaign's description at the moment of publication (i.e., with no user or money activity), making our method applicable for real-time operation on a user browser.

\end{abstract}

%% file: sections/01_introduction.tex
\section{Introduction}
\label{sec:introduction}

Crowdfunding has become a standard means to financially support individuals' needs or ideas, typically through an online campaign appealing to contributions from the community.
What started off as a grassroots movement is now a flourishing industry.
In fact, from \$597M raised worldwide in 2014, to \$17.2B, in North America alone, in 2017, this industry will continue to grow globally~\cite{crowdfundingStatista}.
Over the last decade, the emergence and consolidation of crowdfunding platforms (CFPs) have narrowed the field to a handful of platforms.
Top contenders, such as Kickstarter, FundingCircle, and GoFundMe, have specialized into one of three categories:
investment-based platforms where donors become angel investors in a new enterprise;
reward-based platforms where the backers provide loans with the condition of interest upon repayment;
and donation-based platforms where campaigns are an appeal to charity~\cite{belleflamme2015economics}. 

CFPs and their increasing popularity and fundraising ability for many causes (even recently for coronavirus-related costs~\cite{nyt2020}) inevitably attract malicious actors who take advantage of unsuspecting users (e.g.,~\cite{goFraudMe, nyt2019, theSun2020, ctm2020}).
Immediate access to trusting investors and their funds make these platforms particularly attractive to malicious activity. 
Some platforms allow fund disbursements to happen immediately following a donation, while others require scheduled intervals (\eg weekly), or reaching donation goals.

Furthermore, there is a striking lack of regulation in this space~\cite{fbi2020}.
This void leaves a grey area where crimes are hard to define and difficult to prosecute.
The emergence of few highly publicized cases of fraud in crowdfunding campaigns further undermines general public confidence in this trust-based enterprise.
Nevertheless, evidence of campaign and fund misuse is scant.
According to GoFundMe, one of the most prominent CFPs, fraudulent campaigns make up less than 0.1\% of all campaigns posted on the site~\cite{gfmfraud}. %\footnote{Interestingly, GoFundMe website was recently updated and removed this \%}.
But even at this ``low'' rate of fraud, which has not be substantiated with transparent reports by GoFundMe or other CFPs, in a billion dollar industry, it can amount to tens of millions in defrauded funds every year.
Given CFPs' major source of revenue are these campaigns, through commissions on new campaigns and on every donation, CFPs are not properly incentivized to detect and stop fraud.
Therefore, the lack of tools quantifying this problem is not surprising, effectively preventing the protection of unsuspected contributors.

In this study, we aim to provide such tools to help combat fraud in donation-based CFPs.
We analyze campaigns in North America created to cover medical expenses, a primary reason for these types of appeals (one in three crowdfunding campaigns~\cite{forbesMedical}).
The urgency and strong emotional content of health-related financial constraints easily attract donor attention and donations.
We are interested in quantifying the prevalence of fraudulent behavior in these campaigns, requiring us to classify campaigns as fraud or not. % and is challenging in itself.

Our goal is to create a machine learning (ML) classifier to distinguish between campaigns that are fraudulent or not, at the moment of their creation, i.e., using only features extracted from campaigns newly published.
To accomplish our task, we collect and annotate over 700 campaigns from all major CFPs (GoFundMe, MightyCause, Fundly, Fundrazr, and Indiegogo) and derive deception cues from both the text and the images provided in each campaign.
Overall, we find that fraud is a small percentage of the crowdfunding ecosystem, but an insidious problem.
It corrodes the trust ecosystem on which these platforms operate on, endangering the support that thousands of people receive year on year.
Our results show that using an ensemble ML classifier that combines both textual and visual cues, we can achieve a Precision of 91.14\%, Recall 90.77\% and AUC 96.01\%, i.e., approximately 41\% improvement over the deception detection abilities of people within the same culture~\cite{Bond1990}.
This work is also the first to incorporate text and images in the analysis of fraud, and we rely on features available immediately after a campaign goes live.
This is a significant step in building a system that is preemptive (e.g., a browser plugin) as opposed to reactive.
We believe our method could help build trust in this ecosystem, by allowing potential donors to vet campaigns before contributing.
Similarly, CFPs could use it to prompt vetting and request additional information from a potentially fraudulent campaign creators before campaigns are made public.

Our contributions with the present study are as follows:
\begin{itemize}
\item We collect a dataset of over 700 crowdfunding campaigns on different health-related topics, including medical, emergency appeals, and memorials, and annotate them for being fraudulent or not.
\item Through the use of NLP techniques, we extract language cues, including emotions and complexity of language, and study their association with fraudulent campaigns.
\item Using convolutional neural networks, we extract characteristics of the images posted with each campaign, including emotions and content displayed, and associate them with fraudulent campaigns.
\item Using the above features (text and image-based), we train supervised classification techniques to perform automatic detection of fraudulent campaigns, and discuss our results.
\item We make the collected and annotated dataset available for other researchers to further investigate this problem on CFPs.
\end{itemize}

%% file: sections/02_relatedWork.tex
\section{Related Work}
\label{Sec:related}

Previous work on financial fraud highlights this task's complexity.
Financial information has typically been used to predict the likelihood of a given transaction being fraudulent.
Primarily, past works build behavioral profiles for each user to compute the likelihood of a new transaction being legitimate~\cite{abbasi2012metafraud,bhattacharyya2011data,cecchini2010detecting,dechow2011predicting,panigrahi2009credit,sanchez2009association,srivastava2008credit}.
Novel models in finance technology have opened new areas of research in this space.

Some works focus on detecting deception online.
\citet{luca2016fake} take a set of reviews identified as fraud or not fraud by the platform Yelp and explore the determinants of fraudulent behavior.
The authors explore how restaurants' engagement in positive and/or negative review fraud (i.e., fake reviews) interacts with reputation and competition, over time. We use insights from these fraudulent reviews to shape our understanding of fraudulent deceptive behavior in CFP campaigns.

In their work on Peer-to-Peer lending,~\citet{xu2015p2p} explore trust relationships between borrowers and lenders.
Funds raised can be used for private affairs (\ie there are no business plans and no milestones to rely on for validation), just like crowdfunding campaigns.
However, in their model, the money is meant to be restituted and the network members build up their reputation over time.
They find that soft descriptions of the borrower, \eg age, gender, race, and appearance, are good predictors of whether a loan will be repaid in time.
Conversely, the physical appearance (gender, race, and attractiveness) of the campaign's creator as understood from their profile picture can be used to predict the trustworthiness and, by extension, the success of a campaign~\cite{lin2013judging,pope2011s,duarte2012trust}.
Again, such features help us better understand trust relationships between funders and requesters.

Similarly to Peer-to-Peer lending, crowdfunding is an online financial tool in the hands of millions.
Both are paradigms that depend on participant trust, and fraud against them ``causes emotional and financial harm to lenders (donors) and great damage to sites (platforms) destroying their reputation"~\cite{xu2015p2p}.
The study of fraud in crowdfunding has been tied to the success of the campaign~\cite{wessel2016emergence}, or to entrepreneurial endeavors.
In fact,~\citet{wessel2016emergence} looked at social capital as a means to influence consumer decision making.
They take 591 campaigns that have been flagged for having fake Facebook likes and find that fake social capital has an overall negative impact on the number of backers of a campaign.

~\citet{siering2016detecting} looked at the problem of deception in crowdfunding campaigns, focusing on investment-based donations, where there is an expectation of a reward and a defined business plan. Importantly, these are business interactions, fundamentally different from the altruistic behavior implied in our campaigns.
Based on linguistic text cues, their model presented in~\cite{siering2016detecting} achieves 75\% accuracy.
We build on textual features and combine them with image features, leading our model to achieve 86\% accuracy.

Finally, and most similar to our work,~\citet{cumming2016disentangling} look at entrepreneurial campaigns (\ie commercial campaigns with pledges and rewards) and try to understand the difference between campaigns labeled as detected fraud, suspected fraud, and not fraud.
Using theories from economics and behavioral sciences they identify four possible markers: characteristics and background of the campaign creator (\ie use of names and further participation in the community), a campaigns' affinity to social media, funding and reward structure in the campaign (\ie the duration of the campaign), and finally, details in the campaign description (\ie clarity of language and veracity).
They test their markers in a dataset of 207 fraud cases from two major crowdfunding portals (Kickstarter and Indiegogo) and find that fraudulent campaigns can be described as having longer periods of collection, no Facebook page associated to it, and campaign creators with comparatively less time in the crowdfunding community.
Even though relevant to our work, there are several primary differences with~\citet{cumming2016disentangling}: 1) the type and incentive behind the campaigns (entrepreneurial vs. charitable donations for health problems), 2) when the funds are available (the full amount must be raised vs. immediately available), 3) the tone and the content of the text in the campaigns. 
Indeed, while we do borrow their insights on the relationship between fraud and the simplicity of text, the problem we are addressing is different.

% this work and the work we present. First, the nature of the campaigns: while they focus on entrepreneurial campaigns, the focus of this work is charity based donations. The impact in terms of research is that for example, our campaigns are not time-limited and donors need not raise the full amount before they have access to funds. For entrepreneurial campaigns, the goal is to tap into a wide investor pool, whereas in a donation-based campaign the personal network of the beneficiary (or the creator) is more likely to determine it's success. In terms of the text of the campaign, entrepreneurial campaigns are characterized by fact-oriented descriptions which are normally a few thousand words~\cite{cumming2015crowdfunding}. By contrast, perhaps tied to a network of known relations, donation-based campaigns vary widely in terms of the content, tone, and description of the problem at hand. 
%While we do borrow their insights on the relationship between fraud and the simplicity of text, the problem we are addressing is different.

%\bp{add Yelp paper, check if more on deception}\\

%% file: sections/03_methodology.tex
\section{Crowdfunding Campaign Data Collection and Annotation}\label{sec:dataset}

\begin{table*}
\centering
\caption{Data fields available across crowdfunding sites. These fields were available during the annotation process.
The automated ML analysis focuses on the campaign description, including text and image content.
Note that many fields are variable (e.g., money raised, donors, etc.) and are not available to the ML process for use when the campaign is first published.}
\label{table:campaign_feats}
\begin{tabularx}{\textwidth}{lX}
\toprule
\textbf{Feature}		&	\textbf{Description}\\ \midrule
Creation Date		&	Timestamp of the public release date.\\ 
Campaign Duration	&	For time-limited campaigns, the difference between release and close date. Otherwise, it reflects the time elapsed between release and the time of scrapping.\\
Campaign Status	&	Reflects whether the campaign is open for donations or not (boolean).\\
Title				&	Title of the campaign.\\
Created by		&	Campaigns may be a direct appeal or an appeal made on behalf of another. This field contains a reference to the owner of the campaign who might be different from the beneficiary.\\
Description		&	A narrative of the cause for which the campaign is launched (and updates to the story).\\
Category			&	The general classification of the campaign (\eg Memorial, Health, Emergencies, etc.).\\
Fundraising Goal	&	The total amount of money the creator hopes to raise.\\
Money Raised		&	The money that has been raised by the campaign to date.\\
Number of Donors	&	The number of individual contributors to the cause.\\
Donation Amount	&	The individual amounts from each of the contributions.\\
Social Media		&	The number of likes or shares that the campaign has received over social media.\\
Geo-Tag			&	The location from which the campaign was launched.\\
\bottomrule
\end{tabularx}
\end{table*}

\subsection{Background on CFPs}

Crowdfunding sites are designed to help people connect funding requests with benefactors.
While each CFP is different, they all provide a search engine to find specific campaigns and a classification system that allows visitors to find campaigns that may be relevant to their interests.
In this paper, we looked at campaigns from the top five crowdfunding platforms online: Indiegogo, GoFundMe, MightyCause, Fundrazr, and Fundly.
The information displayed per campaign varies depending on the platform and on the creator of the campaign.
Table~\ref{table:campaign_feats} summarizes the fields and descriptions that are common across all sites.

Depending on the type of campaign and hosting platform, funds raised are delivered either to the campaign creator or its beneficiary.
Typically, entrepreneurial campaigns require a goal to be met before funds are released (not reaching a funding goal results in contributions being returned to investors), whereas in charitable projects, there is no limit as to how soon or often any funds are withdrawn from the campaign.
In terms of revenue, \textit{gofundme.com}, the most prominent CFP, states that they charge a percentage of the transaction as fees, plus a fixed amount per donation~\cite{gfm_pricing}.

Finally, CFPs are aware of the risk of fraud and some offer a guarantee: any member that made a donation to a fraudulent campaign is entitled to a refund by the CFP.
The reimbursement, however, must be requested by the donor after an internal investigation reveals the campaign to be fraudulent.
When a campaign is reported as suspicious, the CFP will send a request for information to the creator of the campaign.
Following the initial report, continued suspicious behavior might result in a campaign being deactivated, or altogether removed.
A deactivated campaign will show the title, primary image and total funds raised.
A removed campaign will result in a redirect to the CFPs main website.
A missing or deactivated campaign, however, is not always an indication of fraud.
For example, a campaign created to raise funds outside a CFP's ``donation cover area'' results in the campaign being removed from the platform.
Alternatively, campaign creators might close a campaign if the fundraising goal has been met, or an event has passed.

\subsection{Defining Fraud}

%
%\begin{table}
%\centering
%\caption{Categories of fraudulent campaigns, as defined across different CFP websites.}
%\label{table:fraud_types}
%\begin{tabularx}{\columnwidth}{lX}
%\toprule
%\textbf{Category}	&	\textbf{Description}\\ \midrule
%Embezzlement		&	Real event and victim, but the funds were never delivered.\\
%Elaborate Fraud	&	Successful campaigns. Events can be caused by the fraudster or made up. \\
%Simple Fraud		&	Unconvincing fraud. Most people would think it is suspicious.\\
%Opportunist		&	Individuals that take recent events and exploit them for financial gain.\\
%Copycat			&	The circumstances are real but the victim / beneficiary have been %changed.\\
%Real				&	The campaign creator thought it to be real (they were mislead).\\
%\bottomrule
%\end{tabularx}
%\end{table}

%% Criminal charges:
% theft by swindle, larceny, wire fraud, child endangerement, failure to make required distribution of funds, mail fraud,  felony grand theft,  false pretense or running a con game, felonious assault, endangering children and telecommunications fraud, money laundering.

Fraud is generally defined as a misrepresentation of an existing fact, made from one person to another, with knowledge of its falsity and for the purpose of inducing the other to act~\cite{fraudDef}.
One requirement of fraud is therefore deception, as it requires the perpetrator to convince the victim that their (false) statement is true.
It also results in damages to the victim and is, most importantly, a criminal offense\footnote{To date, crowdfunding fraud cases have been prosecuted as theft by swindle, larceny, wire fraud, child endangerement, failure to make required distribution of funds, mail fraud, and felony grand theft among others}.

First, we must recognize that fraud is an umbrella term used to define a range of behaviors including embezzlement where (legitimatelly acquired) funds are missappropriated; opportunist fraud where a real story draws criminals to fabricate association to the people/event; or complete fiction in both events and associations, among others. In this work we are attempting to automate the process of recognizing cues available at the time of publication of a crowdfunding campaign where the creator of the campaign is aware of the falsehood of the claims in the campaign. We refer to these campaigns as \textit{fake}.
As an example, opportunist campaigns are \textit{fake}: the creator of the campaign has limited information which can be reflected in his writing style and choice of picture\footnote{In contrast, fraud by emblezzement is \textit{not-fake}}.

In the results we present, our priority is to minimize the number of false positives (\ie real campaigns mislabeled as fraud). However, it is not possible to completely eliminate type II errors (\ie fraud campaigns that were misclassified as real).
Cases like embezzlement where the people, events, and description are real and with the appropriate level of detail but, where the funds were never delivered to the rightful recipient cannot be identified before the decision to commit a crime has been carried out. 
Therefore, the results we present should be understood as a lower bound of the number of cases to be expected in the wild. 
This work, however, is an improvement upon the current state of the art in detection of these types of campaigns where the determination of \textit{fraud} is delegated to the CFPs, and the individual contributors are left on their own devices and judgement to decide if a campaign is fraudulent or not.

\subsection{Datasets}
\label{sec:data}

We have two main sources of data: a set of campaigns that have been confirmed\footnote{Confirmed means either a conviction following a criminal indictment or a condemnation from the victims (more in Section~\ref{gfraudm}).} as fraud and collected from GoFraudMe~\cite{goFraudMe} (we will refer to these as $set\ A$), and two sets of manually annotated campaigns collected from different CFPs ($sets\ B$ and $C$).

\subsubsection{Labeled data from GoFraudMe.com (set A)} \label{gfraudm}
The goal of this website, maintained by an investigative journalist, is to expose fraudulent cases in the GoFundMe platform.
The site serves the dual purpose of holding the CFP accountable for fraudulent campaigns and presenting, preserving, and publicizing the evidence that led to the characterization of fraud.
The site holds 192 confirmed cases of fraud that were shared with us by the website curator.
The process that leads to the inclusion of a campaign in the website varies greatly, but each is accompanied by a narrative that presents the inconsistencies that led to the declaration of fraud. 
Some campaigns have the guarantee of a guilty verdict following legal criminal proceedings, whereas others have been denounced by beneficiaries and supported by their community.
Some of the cases presented, typically those that follow from events reported in various news platforms, give rise to several fraudulent campaigns.
For some cases, there is an archived version of the campaign that was used to collect money with a link to the rightful beneficiary.
For others, there is a timeline following the investigation, that led to criminal charges (and subsequent conviction if it is available).

\subsubsection{Annotated Data (sets B \& C)}
In addition to the labeled campaigns collected from $set\ A$, we created two manually annotated datasets.
$Set\ B$: 191 campaigns from the Medical category in GoFundMe.com.
$Set\ C$: 350 campaigns from different CFPs that were directly related to organ transplants.
$Sets\ B$ and $C$ were manually annotated following the methodology described in the Section~\ref{sec:annotation}.
$Set\ C$ is a random sample of 350 campaigns related to organ transplants collected in January 2019 from the top 5 CFPs: Indiegogo, GoFundMe, MightyCause, Fundrazr, and Fundly.
Both $B$ and $C$ sets were collected through automated crawlers written in python.
From each campaign, we collected the features in Table~\ref{table:campaign_feats} as well as all comments, pictures, and individual donations.
Each campaign was visited in the order presented by the CFP's search engine.
While some CFPs provided APIs to connect with their database, the data fields were collected, for the most part, through the corresponding elements in HTML.

\subsection{Campaign Annotation}
\label{sec:annotation}

%%%%%%%%%%%%%%%%%%%%%%%%%%%%%%%%%%%%%%%%%%%%%%%%%%%%%%%%%%%%%
%We also give details on how we annotated the data.
%
%the annotation effort, the alignment of labels, etc.
%
%Give any interesting insights on the labeling process..
%
%############################################################
%############################################################

\begin{table}
\centering
\caption{Breakdown of the labels for the 733 annotated campaigns. All campaigns had a textual description and many had several images as part of their appeal.}
\label{table:annotation_total}
\begin{tabularx}{\columnwidth}{clXX}
\toprule
\textbf{Score}	&	\textbf{Label}		&	\textbf{Text}	&	\textbf{Images}\\ \midrule
0			&	invalid				    &	93			&	117			\\
1			&	fraud					&	141			&	138			\\ 
2			&	probably fraud			&	26			&	71			\\
3			&	unknown				    &	105			&	123			\\
4			&	probably not-fraud		&	78			&	141			\\
5			&	not-fraud				&	290			&	517			\\
\bottomrule
\end{tabularx}
\end{table}

Combined, $sets\ A,\ B$ and $C$ make up the ground truth in the study. $Sets\ A$ and $B$ are inversely balanced in the sense that one provides mostly examples of fraud  and the other mostly not-fraud\footnote{From our findings, the prevalence of fraud in a random sample of medical campaigns is approximately 10\%, in contrast to the 0.1\% claimed by CFPs. Therefore most of the campaigns we looked at in $Set \ B$ were not-fraud.}.
$Set\ C$ was created as a means to augment the number of campaigns in the study.
All campaigns were manually annotated using the scale proposed in Table~\ref{table:annotation_total}, where (1) indicates certainty of fraud and (5) certainty of not-fraud.
During manual annotation, two expert annotators developed guidelines to determine the label of each campaign. 
The considerations in the guidelines included:
\begin{itemize}
	\item A personal (offline) knowledge of the circumstances that led to the appeal as evidenced in the support messages posted to the campaign. Knowledge is reflected (but not limited to) in having met the beneficiary (or having first hand knowledge of the circumstances), participating in offline fundraising activities, or familial relationships between donors. 
	\item A sense of closure to each campaign, particularly those that have been open for donations for several years.
	\item Coherency between the description, support documents, pictures, fundraising goal, donors, and level of detail.
	\item Participation of the creator in other campaigns.
	\item Reverse search of pictures and text diplayed in the campaign leading to unrelated results in the web.
	\item Evidence of contradictory information.
	\item Overwhelming lack of engagement of campaign donors.  
\end{itemize}

\begin{figure*}[t]
\centering
\subfloat[Text]{\includegraphics[width=.45\linewidth]{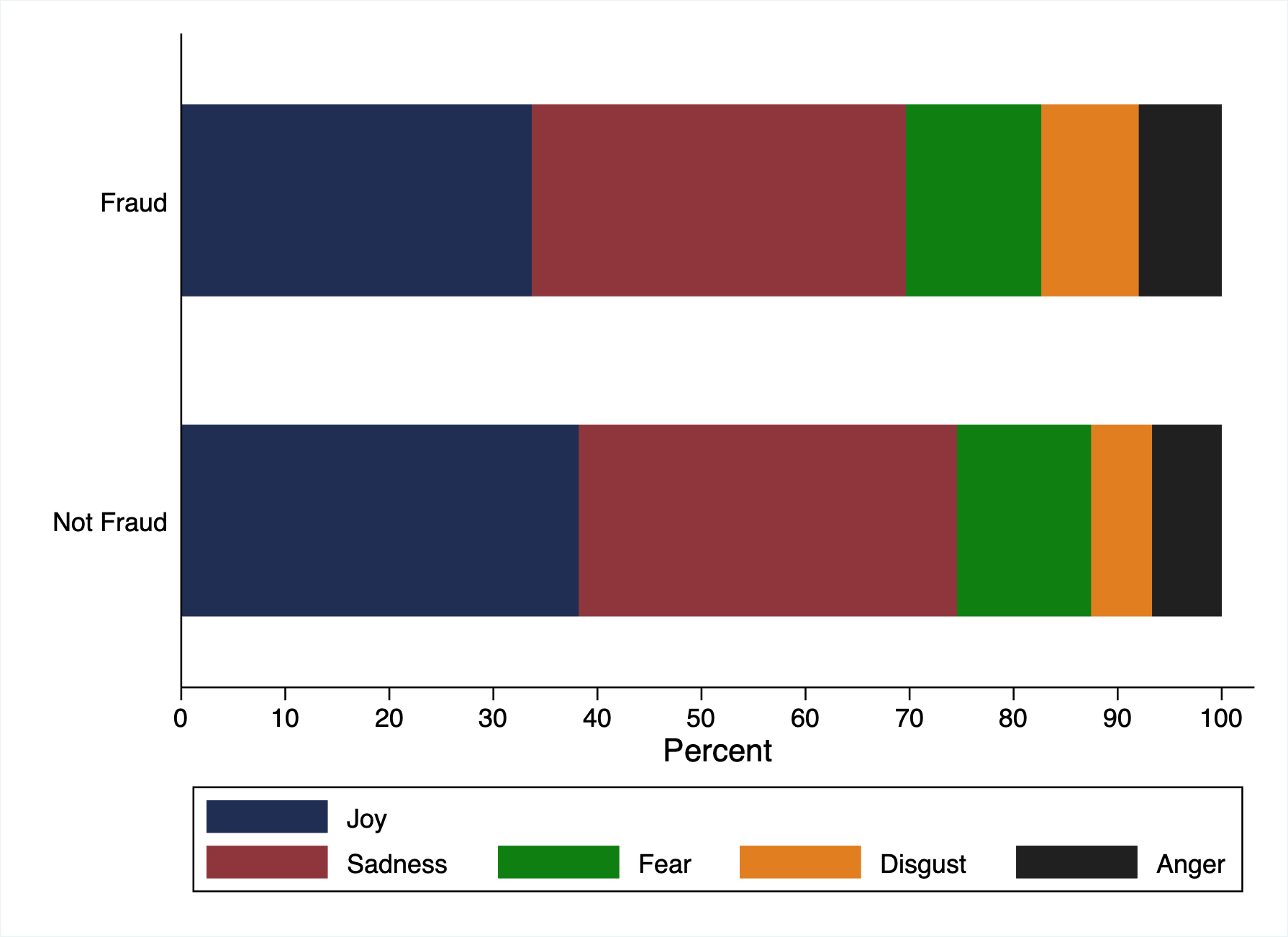}}
\quad %add desired spacing between images, e. g. ~, \quad, \qquad etc25\textwidth} %(or a blank line to force the subfigure onto a new line)
\subfloat[Images]{\includegraphics[width=.45\linewidth]{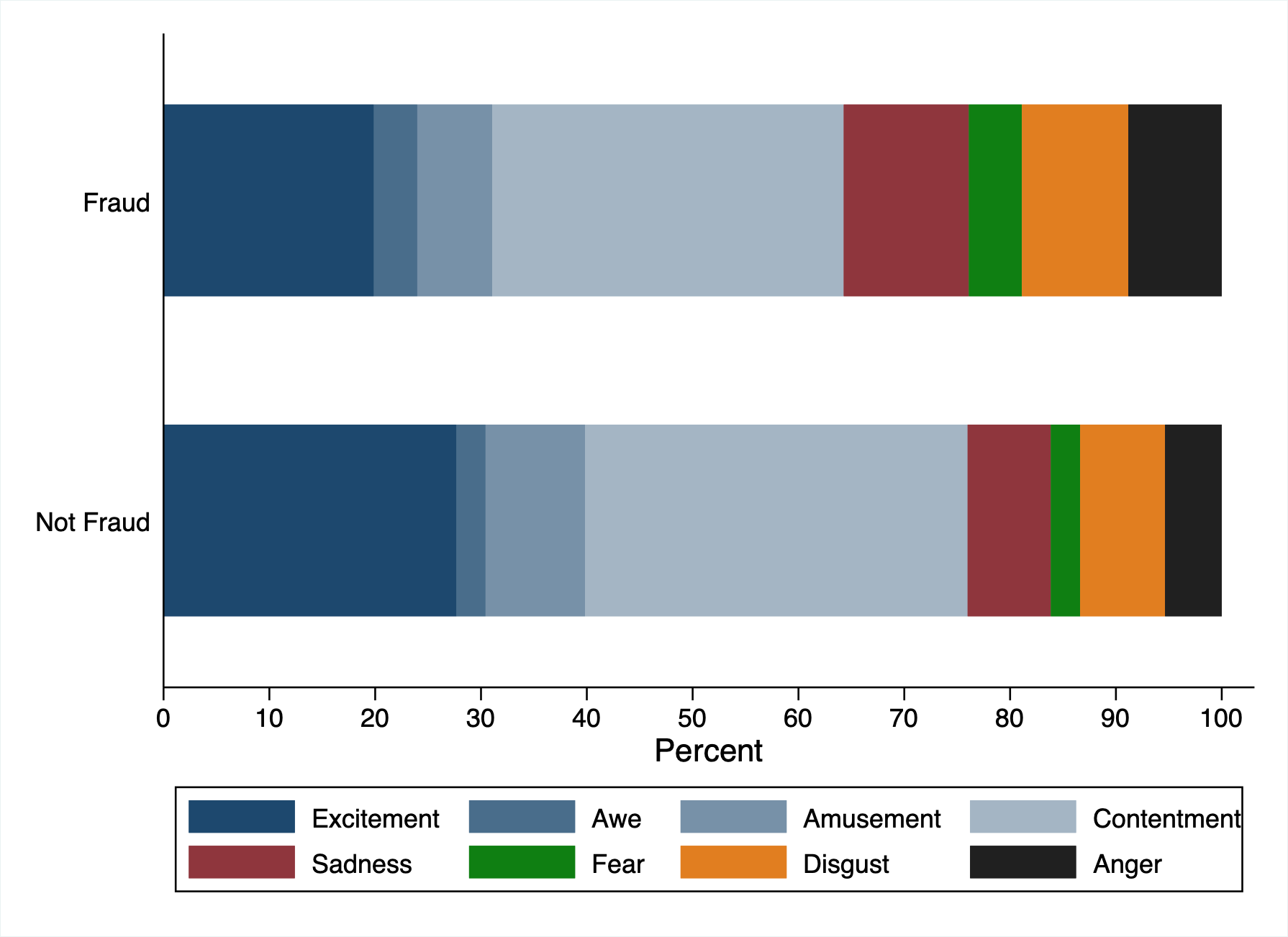}}
\caption{Comparing the emotions displayed in the text and image of the campaign.
}
\label{fig:emotion}
\end{figure*}

The label of fraud assigned by the annotators was independent of the features engineered for automated detection. The annotators relied on semantic interpretation.
The features used in the models are stylistic markers of the textual description and the images in the campaign.
To this extent, we are reasonably certain that while the label of fraud might be incomplete (\ie it will not capture all categories of fraud), it is correct.
Ultimately, we considered 704 campaigns in the study, with some campaigns removed from the dataset because the content was no longer accessible, the text was in multiple languages, or there were too few characters to compute any of the text-based features.

Inter-annotator reliability (or consistency) refers to the validity of the variable being measured~\cite{mchugh2012interrater}.
In this project, we had a structured subjective task, where we iteratively refined and applied a set of guidelines that define fraud to each of the campaigns reviewed.
At the end of the first iteration, annotators reconciled the labels and revised the guidelines.
At the end of the second round of annotations, and for campaigns with contradictory labels, the final decision was agreed by discussion and consensus between annotators.
We measured the consistency across annotators for each iteration using Cohen's Kappa ($\kappa$)~\cite{cohen1960coefficient}.
We applied the interpretation scale proposed by Landis and Koch~\cite{landis1977application}, where values between 0.6 and 0.8 are considered to reflect a substantial agreement between annotators.
At the end of the first iteration, $\kappa$ was found to be $0.451$, reflecting only moderate agreement between annotators.
After revising the guidelines, at the end of the second round of annotations, $\kappa  = 0.675$.
Ultimately, the values used as labels for classification were considered of binary form, \ie 1 for fraud and 0 for not-fraud, that reflect consensus across annotators (\ie a dataset with $\kappa = 1$).

%% file: sections/04_texts.tex
\section{Campaign Fraud: Textual Cues}
\label{sec:text}

The description of the campaign is the best line of communication between the campaign creator or beneficiary and any potential donors.
Therefore, it is the first place where a malicious actor might leave traces of deception.
In this section, we present five different areas where automated analysis might find quantitative evidence of deception.
We also present some preliminary analysis of the variable categories with respect to our classification variable: \textit{fraud}.

\subsection{Feature Extraction}

\subsubsection{Sentiment Analysis.}
We extract the sentiment and tone expressed in the text for further analysis using IBM services~\cite{IBM}.
The sentiment is computed as a probability across five basic emotions: sadness, joy, fear, disgust, and anger.
Complementary to emotions, the text's tone can also express a campaign's intent.
We analyze confidence scores for seven possible tones: frustration, satisfaction, excitement, politeness, impoliteness, sadness, and sympathy.

\subsubsection{Complexity and Language Choice.}
%Domain specific models to understand complex unseen data.
The need for appeal to a more general population can lead fake campaign creators to adapt (or carefully select) the language used.
Simpler language and shorter sentences can appeal to the emotions of the reader and, therefore, be more successful.
To check the language complexity of the document and word choice, we look at a series of readability scores (\eg automated readability index, Dale-Chall Formula, etc.) and language features (\eg function words, personal pronouns, average syllables per word, total number of characters, etc.)~\cite{cumming2016disentangling,wessel2016emergence}.

\subsubsection{Named-Entity Recognition}
Named-Entity recognition is the process of identifying named entities (e.g. proper nouns, numeric entities, currencies) in unstructured text and assigning them to a finite set of categories. In this project, we relied on spaCy~\cite{honnibal2017spacy} a tool released for Python which identifies 18 types of entities in text.
SpaCy models are based on convolutional neural networks built with pre-trained vectors which give an accuracy of 86.42\%.

\subsubsection{Form of the text}
The next group of features we considered was the visual structure of the text. For the entire textual dataset we captured the form of each word: whether the letters were all lower-case, all upper-case, the number of emojis on the text, the number of words with exclamation mark, the words with apostrophes, and many others. We generated a vector with 255 descriptors and evaluated the text in each campaign against the features.

\subsubsection{Word Importance}
Lastly, we considered the numerical vectorial representation of the text given by tf-idf.
This method, similar to a bag-of-words approach, highlights content similarity between different documents. As with the other textual features, we compute word importance on the text included in the campaign description. Ultimately, this description is the primary method of communication between campaign creator and potential donors. While the success of a campaign is mostly determined by the strength of a community and their participation in the system, a good story may persuade chance visitors to donate to the cause.

\subsection{Exploring the Data: Text-Based Features}

%%%%%%%%%%%%%%%%
%\begin{figure*}
%\centering
%\subfloat[Fraud]{\includegraphics[width=.45\linewidth]{figures/word_cloud_fraud}}
%\quad %add desired spacing between images, e. g. ~, \quad, \qquad etc25\textwidth} %%(or a blank line to force the subfigure onto a new line)
%\subfloat[Not-Fraud]{\includegraphics[width=.45\linewidth]{figures/word_cloud_notFraud}}
%\caption{Word importance across campaign descriptions.}
%\label{fig:text_wordCloud}
%\end{figure*}
%%%%%%%%%%%%%%%%%

\begin{figure} [th]
\centering
    \includegraphics[width=\linewidth]{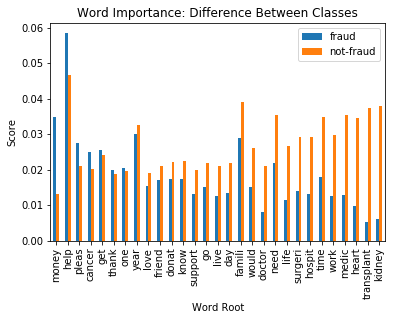}
    \caption{Word importance in campaign description across fraudulent and not-fraudulent campaigns. From left to right, the words are arranged in order of decreasing difference between the two classes.}
\label{fig:text_wordCloud}
\end{figure}

Fig~\ref{fig:emotion}(a) presents the sentiment analysis for the text of the campaigns.
%Joy (shown in blue) is the only positive emotion analyzed.
Each bar is the aggregation of emotions for the label indicated.
From the figure, we see that as emotions, joy and disgust contain valuable information in the separation of the binary variable.
Interesting is also the balance between the positive and negative emotions in each campaign.
This figure shows that, as emotions in the text, campaigns that are not-fraud display more joy and less disgust than campaigns that are fraud.
Almost as if the narrator is making an effort to present their friend or relative (\ie the beneficiary) as they were; then, present the (presumably negative) reason for creating the campaign.

One of the most interesting results we found is evidenced in Figure~\ref{fig:text_wordCloud}.
Word importance for each category shows that while, generally, both sets of campaigns have similar characteristics, fraudulent campaigns are perhaps more desperate in their appeal.
Starting from the left side of x-axis, Figure~\ref{fig:text_wordCloud} shows that the words \textit{money, help, please, cancer}, and \textit{get} are more prevalent in fraudulent campaigns, whereas not-fraud descriptions will emphasize words like \textit{kidney, transplant, heart, medic(al)}, and \textit{work} (right side of x-axis).
In general, legitimate campaigns are more descriptive, being open about the circumstances in making their appeal.

\subsection{Significance: Reducing Dimensionality}
Combined, the five types of text-based analysis result in 8,341 features extracted from the description provided with the campaign, but several of them can be sparse (e.g., TFIDF) and others may not prove to be so helpful in detecting fraud.
%Therefore, we expect that in our task of detecting fraud, some of the features we extracted will be more relevant than others.
Therefore, the final step in our pre-processing of the data is to analyze each feature with respect to the variable we are interested in, and filter out features not useful. 
We start by making no assumptions about the distribution of our random variables and choose the non-parametric, two-sample KS test to check whether the difference between the distributions of the fraud and not-fraud data for each feature are significant at level $a=0.05$.
This testing removed features that were not different, and reduced the space to 71 variables from all five textual analysis categories.
Ultimately, in this paper, any result computed with text-based features includes only the 71 KS significant features\footnote{We ran a similar analysis using the t-test which assumes that the random variable is normally distributed and achieved similar performance with the classifier.}.

%% file: sections/05_images.tex
\section{Campaign Fraud: Visual Cues}
\label{sec:images}

Though the text is the primary means of information, pictures provide the often essential supporting details of the claim.
As with Section~\ref{sec:text}, in this section we present the rational for the features derived from images and the preliminary results of the analysis of the data collected.  

%%%%%%%%%%%%%%%%
%\begin{figure*}[t]
%\centering
%\subfloat[Fraud]{\includegraphics[width=.45\linewidth]{figures/object_cloud_fraud}}
%\quad %add desired spacing between images, e. g. ~, \quad, \qquad etc25\textwidth} %%(or a blank line to force the subfigure onto a new line)
%\subfloat[Not-Fraud]{\includegraphics[width=.45\linewidth]{figures/object_cloud_notFraud}}
%\caption{Most prevalent objects in images across Fraudulent and Not-Fraudulent campaigns.}
%\label{fig:objectAnalysis}
%\end{figure*}
%%%%%%%%%%%%%%%

\begin{figure}
\centering
    \includegraphics[width=\linewidth]{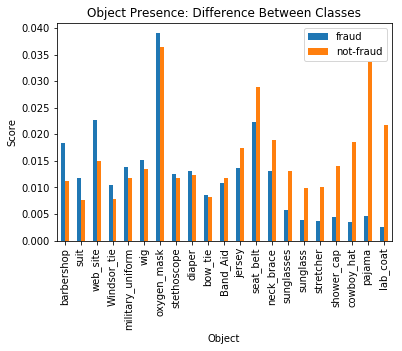}
    \caption{Object prevalence in images across fraudulent and not-fraudulent campaigns. From left to right, the objects are arranged in order of decreasing difference between the two classes.}
\label{fig:objectAnalysis}
\end{figure}

\subsection{Feature Extraction}

\subsubsection{Emotion Representation.}
Psychological studies show that images, as a form of visual stimuli, can be used to induce human emotion~\cite{joshi2011aesthetics}.
Visual emotion prediction has therefore attracted much interest from the computer vision community---framed as a multiclass classification problem using image-emotion pairs as input-output tuples for learning.
Motivated by the foregoing successes for visual emotion prediction in transfer learning, we repurposed a ResNet-152~\cite{he2016deep} a convolutional network pre-trained on the ImageNet dataset~\cite{krizhevsky2012imagenet} containing 1.2 million images of 1000 diverse object categories.
The fine-tuning was performed by replacing the original 1000-way fully connected classification layer with a newly initialized layer consisting of 8 neurons that correspond to the emotion categories of interest.
As defined in~\cite{zhao2014exploring,you2016building}, the eight categories were as follows: amusement, anger, awe, contentment, disgust, excitement, fear, and sadness.

To fine-tune the model, we utilized the Flickr and Instagram (FI) dataset~\cite{you2016building} of 23k images, where each image is labeled as evoking one of the eight emotions based on a majority vote between five Amazon Mechanical Turk workers.
We used 90\% of the images for training and the remainder for validation.
During pre-processing, each image was resized to $256\times256\times3$ and standardized (per channel) based on the original ImageNet training data statistics.
We used 100 epochs, to minimize a negative log-likelihood loss, with stochastic gradient descent, using an initial learning rate of 0.1, momentum 0.9, and a batch size of 128.
The learning rate was multiplied by a factor 0.1 at epochs 30, 60 and 90.
We performed data augmentation by randomly cropping $224\times224\times3$ image patches, which is the resolution accepted by ResNet-152.
During fine-tuning, all layers except the classification layer were frozen.
The final model accuracy on the validation data was 73.9\%, where predictions are based on central $224\times224\times3$ crops.

The semantic evidence over the eight emotions, in the form of logits (unnormalized log probabilities), can then be extracted for the crowdfunding images.
Each image was resized such that its shortest side was 256 pixels and then a central crop was extracted of size $224\times224\times3$.
Semantic emotion category representations were then extracted from the classification layer.

\subsubsection{Appearance and Semantic Representations.}

Again, with the help of a pre-trained ResNet-152 model, trained on the ImageNet dataset, we extracted appearance representations and semantic representations of each of the images present in the campaigns. For pre-processing, each crowdfunding image was resized such that its shortest side was 256 pixels and then a central crop extracted of size $224$$\times$$224$$\times3$. We standardized each image (per channel) based on the original ImageNet training data statistics.

The appearance representations is meant to quantize the picture itself by generating a vector of descriptors from the penultimate layer of the network. These features ( $\in\mathbb{R}^{2048}$) provide a description of each image where the fields, automatically learnt by the network can be, \eg the dominant color, the texture of the edges of a segment, a (lower level) object --- \eg an eye, among others.
In contrast, the semantic representation expresses the logit presence of pre-determined objects in each image.
The vector ($\in\mathbb{R}^{1000}$) is extracted from the classification layer over the 1000 ImageNet classes.

Each representation is useful since convolutional neural networks are known to implicitly learn a level of correspondence between particular objects~\cite{zeiler2014visualizing}.
Moreover, the representations invariably outperform their hand-engineered counter-parts~\cite{chatfield2014return}.

Finally, we consider the number of faces present in the image as a possible distinguishing factor between \textit{fraud} and \textit{not-fraud} campaigns. We extract this feature using the dlib~\cite{king2009dlib} HOG-based face detector and estimate the number of faces present per image.

\subsection{Exploring the Data: Image-Based Features}

\begin{figure}
\centering
	\includegraphics[width=0.85\linewidth]{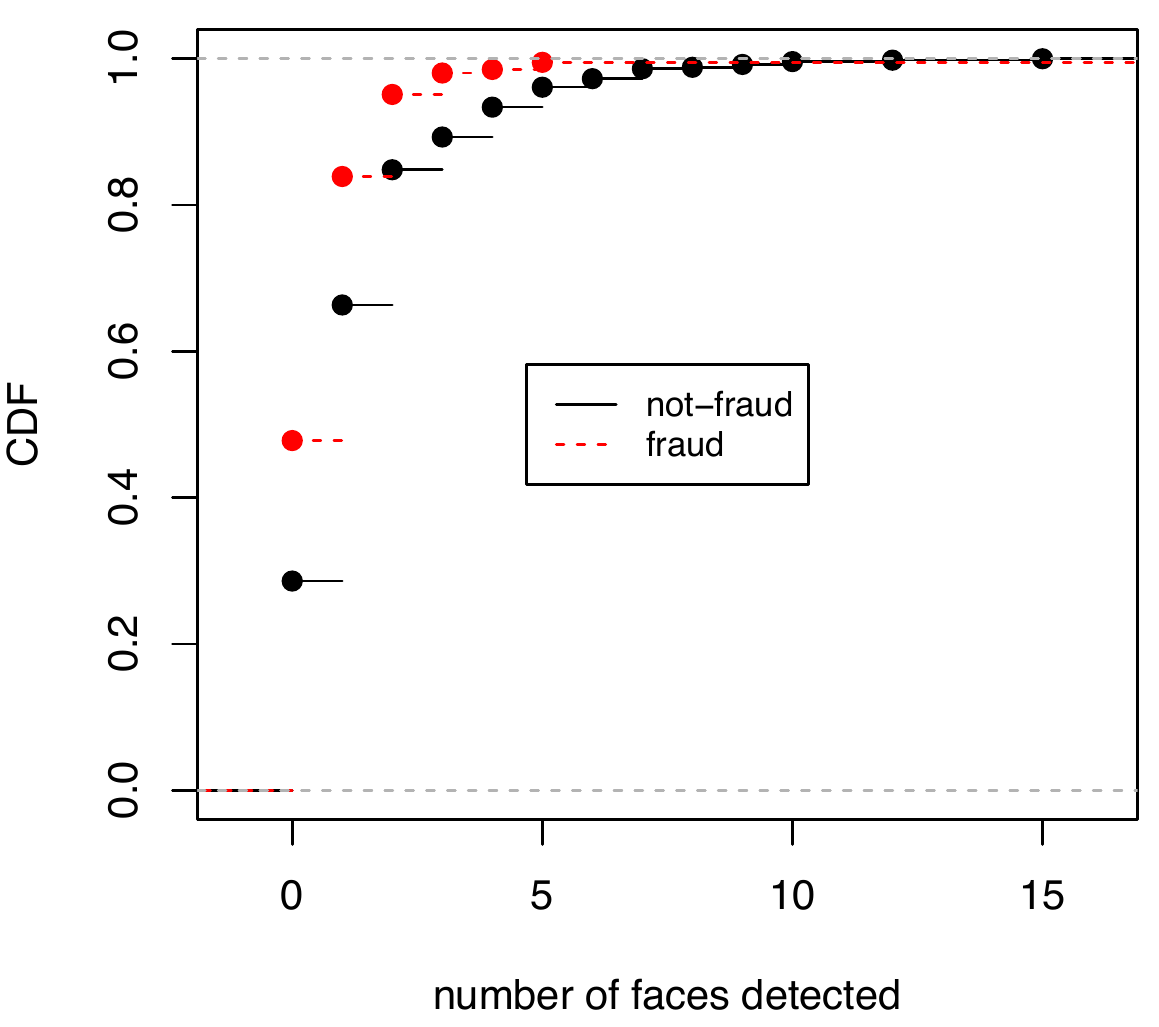}
	\caption{Number of faces detected in images across fraudulent and not-fraudulent campaigns.}
\label{fig:faces-detected}
\end{figure}

In our analysis of emotion in images we found that, as compared to the text (in Figure~\ref{fig:emotion}(a)), there is a greater imbalance between positive emotions and sadness. Figure~\ref{fig:emotion}(b) shows the positive emotions as shades of blue and other emotions as indicated in the legend.
Similar to the text, not-fraud campaigns display more positive emotions and proportionally less anger and fear through their images. 

In our analysis of objects present in each image (Figure~\ref{fig:objectAnalysis}), we find that not-fraud campaigns have a stronger presence of objects that are associated with hospital stays (as evidenced by the presence of objects like \textit{lab coats, pajamas, stretchers}, and \textit{neck braces}) though the same categories are found to a lesser degree in the fraudulent campaigns.
On the other hand, fraudulent campaigns appear to include images with objects or concepts that are more casual in nature, such as \textit{barbershop, suit, tie, uniform}, which may not fit the context of CFP campaigns launched for medical-related problems.

Compared to the results in Figure~\ref{fig:text_wordCloud}, the signal revealed by the images is not as strong as the one contained in the text, as the separation is not so clear.
The difference between text and images can be explained by considering that CFPs provide, at times, specific instructions regarding the types of images to include.
For example, one such instruction is to include a picture of the fundraising organizer and the person in need looking happy.
Not only does this homogenize the type of images used in the fundraisers, it also provides a clear guidebook for potentially fraudulent campaigns, hence diminishing the predictive power of images, in general, and the objects identified in those images, in particular. 

We also analyze the number of faces detected in the images of the two types of CFP campaigns, in Figure~\ref{fig:faces-detected}.
Even though the number of faces in the extreme cases (e.g., above 10 faces detected) follows similar distribution in the two classes, we notice that for the majority of non-fraudulent campaigns, they tend to include images with more faces than fraudulent campaigns.
Interestingly, the median for both classes is 1, but the mean for non-fraudulent campaigns is 1.488 and for fraudulent is 0.8341, which means it is more common to include images with at least one face in the non-fraudulent campaigns.

\subsection{Significance: Reducing Dimensionality}
Combined, the visual cues amount to 3,057 features. As was the case with textual features, we expect the semantic representation of each image to be sparse and some features to be more discriminative than others with regards to our target variable.
As was the case with the text-based features, we used the KS-test to determine the significance of each descriptor.
The result was a vector of 501 features with representatives from all categories: emotion, appearance, semantics, and number of faces.
The classification models contained only these 501 features in all types of image analysis.

%% file: sections/06_modeling.tex
\section{Automated Detection of Crowdfunding Fraud}
\label{sec:results}

Next, we present our effort to train a machine learning (ML) classifier to automatically detect fraudulent campaigns using various features discussed in the previous sections.

\begin{table}
\centering
\caption{Average precision and accuracy for different classification algorithms, for the \textit{Label II} setup and with text-based features considered (st. deviation shown in parenthesis).}
\label{table:text_classifiers_all}
\begin{tabular}{lccc} %{\columnwidth}{IXXX}
\toprule
\textbf{Classifier}	& \textbf{Accuracy}	& \textbf{F1-Score}& \textbf{AUC}\\ \midrule
SVM				    & 0.6223 (0.078)	& 0.6204 (0.079)   & 0.6223 (0.076)\\
\textit{k}-NN		& 0.6252 (0.077)	& 0.6116 (0.083)   & 0.6252 (0.070)\\
Naive-Bayes		    & 0.7980 (0.062)	& 0.7967 (0.063)   & 0.7980 (0.062)\\
AdaBoost			& 0.8061 (0.063)	& 0.8060 (0.063)   & 0.8061 (0.063)\\
Decision Tree		& 0.8130 (0.062)	& 0.8129 (0.063)   & 0.8130 (0.062)\\ 
Random Forest		& 0.8368 (0.059)	& 0.8367 (0.059)   & 0.8368 (0.059)\\
MLP				    & 0.8553 (0.050) 	& 0.8544 (0.050)   & 0.9252 (0.040)\\
\bottomrule
\end{tabular}
\end{table}

\subsection{Experimental Setup}

\subsubsection{Fraud Scale Grouping.}
% overall dataset, testing the labels together
The fraud scale presented in Table~\ref{table:annotation_total} can be combined in different ways to generate the overall label of \textit{fraud}.
In our first experimental setup, we use the union of campaigns with scores \{1,2\} as \textit{fraud}, scores \{4,5\} as \textit{not-fraud}, omitting the campaigns with score 3, and denote this setup as \textit{Label I}.
In the second experimental setup, we define as \textit{fraud} exclusively the campaigns with scores of \{1\}, and \textit{not-fraud} the campaigns with scores of \{5\}, omitting the other campaigns, and denote this setup as \textit{Label II}.
Practically, in using \textit{Label I}, we prioritize the need to get more observations for the training of the classifier, whereas using \textit{Label II}, we give more importance to the strength of the signal being captured, but in reduced instances.
In our experiments, we observed better performance when minimizing the noise in the signal.
Ultimately, we chose \textit{Label II} for the final results.

\subsubsection{ML Classifiers.}
In choosing a classifier, we need a method that is fast, robust to noise and not prone to overfit the data, thus, allowing the model to be generalizable.
We tested different classical ML methods whose implementation is available in sklearn~\cite{scikit}: Random Forests (RF), AdaBoost, Decision Tree, \textit{k}-NN, Naive-Bayes and Support Vector Machine (SVM), and compare their performance across different metrics. In addition to the classical methods, we also built a multilayer perceptron (MLP) with one hidden layer (followed by a ReLU) of dimensionality equal to its input. Each MLP was trained for 50 epochs using SGD with momentum $0.9$, weight decay \num{5e-4}, a batch size of 1 and initial learning rate of $0.001$.
During training, inputs were corrupted on-the-fly with additive white Gaussian noise ${\sim N(0,\sqrt{0.1})}$.
%We compare all these algorithms and highlight the ones offering a good compromise between the above conditions.

\subsubsection{Performance Metrics.}
For each classifier, we compute five metrics: accuracy, precision, recall, F1-score, and the area under the ROC curve (AUC), which plots the relationship between true positives and false positives at different operating thresholds of the classifier.
%Accuracy is a measure of the number of hits (\ie correctly labeled test observations).
%Precision is a measure of the positive predictive power of the classifier, recall measures the number of true cases missed by the classifier, the F1-score is the harmonic mean of the precision and the recall, and the AUC is a measure of chance where values close to 0.5 make the classifier uninformative.
For these metrics, a perfect classifier would score 1 in all.

%\begin{align*}
%Precision &=\frac{\text{TP}}{\text{TP + FP} } & Recall&=\frac{\text{TP}}{\text{TP + FN}}
%\end{align*}
%\begin{align*}
%F1 &= \frac{2 \cdot precision\cdot recall}{precision+ recall} & Accuracy &=\frac{\text{TP + TN}}{\text{TP + TN + FP+ FN}}   \\
%\end{align*}

\subsubsection{Experiment Iterations.}
Initial attempts at classification showed that the classifiers' results for different metrics were dispersed.
To obtain accurate measures of each model's performance, and following the law of large numbers, we increased the number of iterations and looked at the distribution of results for each model.
For each iteration, we perform a random split of train and test data.
As expected, multiple iterations over the different splits of data yielded different results.
Overall, the mean of normally-distributed classification results can approximate the true value of each metric.
Also, the classes are not balanced and, therefore, we forced the same number of observations for each class by under-sampling the bigger class (\ie a random selection of observations available) while creating a split per iteration.

We perform two experiments: a preliminary one, to test the performance of each feature modality (text vs. images), and then the final one with an ensemble classifier that uses an average of the two preliminary ones.
Results for the classical ML algorithms were computed by executing 2,000 iterations of the classifiers on the available text or image data. For the neural network, we used 1,000 models to obtain the final classification. 

\subsection{Predicting from Different Modalities:\\Text vs. Images}

%%%%
%\nk{Also, we need to argue why not run a RF with just a large number of trees, e.g., 10000}

Tables~\ref{table:text_classifiers_all} and~\ref{table:image_classifiers_all} show the performance of the considered classifiers, using \textit{Label II}, with textual and visual features, respectively.
These results were obtained by running 2,000 iterations and computing all metrics for each model.
As shown in the tables, all classifiers outperform the 50\% random baseline of binary classification implying that the signal separating \textit{fraud} from \textit{not-fraud} is present in the data.
Interestingly, tree-based models such as Decision Tree and Random Forest perform fairly well with  AUC up to 0.84, just under the 0.93 AUC exhibited by neural networks on textual features.
%However, the MLP neural network performs the best with AUC of 0.93 on the text-based features.
We note that textual data alone provide better classification power than images alone (AUC=0.93 vs 0.67). However, the classification performance is improved by combining modalities.

%%%%%################

\begin{table}
\centering
\caption{Average precision and accuracy for different classification algorithms, for the \textit{Label II} setup and with visual features considered (st. deviation shown in parenthesis).}
\label{table:image_classifiers_all}
\begin{tabular}{llll} %{\columnwidth}{IXXX}
\toprule
\textbf{Classifier}	& \textbf{Accuracy}	& \textbf{F1-Score}& \textbf{AUC}\\ \midrule
SVM				    & 0.6590 (0.061)	& 0.6586 (0.061)   & 0.6622 (0.061)\\
\textit{k}-NN		& 0.6350 (0.061)	& 0.6319 (0.062)   & 0.6620 (0.062)\\
Naive-Bayes		    & 0.6584 (0.062)	& 0.6576 (0.062)   & 0.6605 (0.061)\\
AdaBoost			& 0.6423 (0.062)	& 0.6419 (0.062)   & 0.6609 (0.061)\\
Decision Tree		& 0.5701 (0.064)	& 0.5693 (0.065)   & 0.6597 (0.060)\\ 
Random Forest		& 0.6746 (0.061)	& 0.6741 (0.062)   & 0.6787 (0.061)\\
MLP				    & 0.6230 (0.056)	& 0.6167 (0.061)   & 0.6737 (0.063)\\
\bottomrule
\end{tabular}
\end{table}

%%%%%################

\subsection{Automatically Detecting Campaign Fraud}

%%%%%% ABCII Ensemble Compare
\begin{table}
\centering
\caption{Evaluation metrics for the ensemble classifiers using the \textit{Label II} setup (st. deviation shown in parenthesis).}
\label{table:ensemble_ABCII_compare}
\begin{tabular}{lcc}%{\columnwidth}{lXXX}
\toprule
Metric	& \textbf{RF Ensemble} & \textbf{MLP Ensemble} \\ \midrule
Accuracy	& 0.8517 (0.068) & 0.9014(0.034) \\ 
F1-Score	& 0.8517 (0.068) & 0.9013(0.034) \\
AUC		    & 0.8539 (0.068) & 0.9601(0.022) \\
\bottomrule
\end{tabular}
\end{table} 

The models based on text and images show definite separation between the class (\textit{fraud} or \textit{not fraud}).
The next step is to determine whether combining all features of a campaign into the same model provides improvement over treating them separately. Tables~\ref{table:text_classifiers_all} and~\ref{table:image_classifiers_all} show that RF is the best from the classical algorithms, but MLP outperforms RF in textual features.
%However, looking at accuracy, for the textual-based case, the MLP outperforms RF.
Thus, we use both RF and MLP to evaluate the ensemble classifier performance.
Furthermore, Table~\ref{table:annotation_total} showed that information on each campaign varies: some have no images while others have multiple.
We first run the classification task separately for text and images, and then combine results into a single score for each campaign.
As before, we train and test RF with 2K runs, and the MLP on 1K models, on the \textit{Label II} setup. We then run an ablation study to determine whether any of the feature groups (\ie TFIDF, text Sentiment Analysis, Named-Entity Recognition, the Shape of the word, the Readability Index, the Descriptive elements of an image, the Objects present in an image, Emotions triggered by each image, and the number of faces recognized) have a negative interaction and should therefore be removed.  % (strengthening the clarity of the signal).

The results, shown in Table~\ref{table:ensemble_ABCII_compare} indicate that, while the neural network approach was comparable to the classical algorithms in terms of the separate modalities (\ie images and text), there is a clear improvement in all metrics when we combine all features in the same model, with AUC=$0.96$.

\begin{figure}
\centering
    \includegraphics[width=\linewidth]{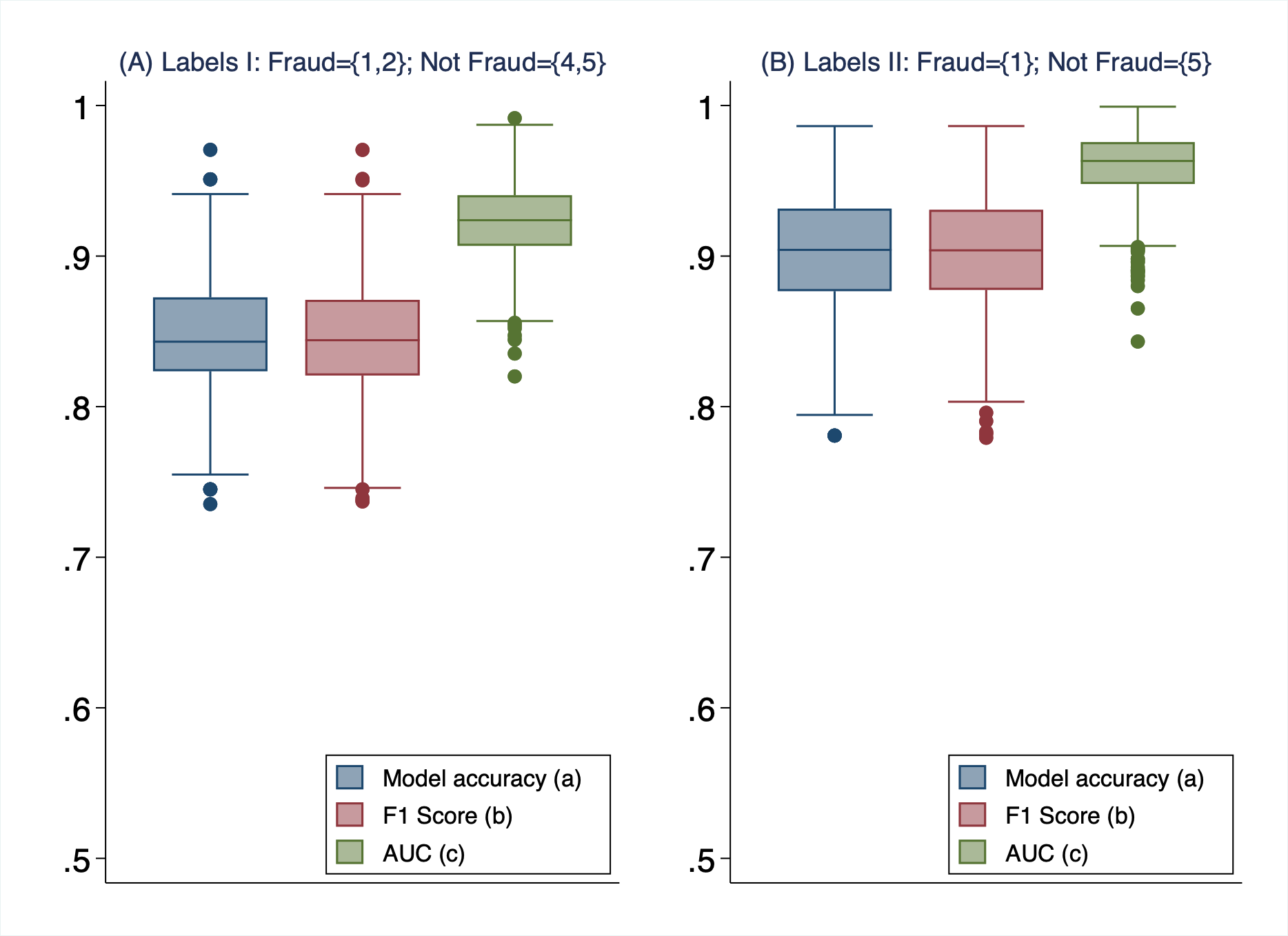}
    \caption{Box diagram of the classification results for 1k models of the neural network (MLP) classifier.}
\label{fig:ABC_ensemble_distribution}
\end{figure}

For completeness, Figure~\ref{fig:ABC_ensemble_distribution} presents the distribution of the individual evaluations of the neural network for the \textit{Label I} and \textit{Label II} setups.
As expected, \textit{Label II} performs better than \textit{Label I}.
Also, the models are not dispersed, and the results are consistently over 80\% of median performance.

\subsection{Clarity: Classifying imperfect data.}
In Section \textit{6.1.1}, we discussed the impact the labels have on the classification output.
Here, we investigate another configuration presented as \textit{Label III}.
This corresponds to the scenario where we train on campaigns with scores of \{1\} for \textit{fraud}, and campaigns with scores of \{5\} for \textit{not-fraud} (i.e., \textit{Label II} setup), and then test this model on campaigns with label scores \{2,4\}, corresponding to \textit{fraud} and \textit{not-fraud}, respectively.
These campaigns were dropped in \textit{Label II} setup, and thus were unseen by the classifier.

In Table~\ref{table:ABC_ensemble_distribution}, we compare the performance of modeling fraud with \textit{Label I}, \textit{Label II} and \textit{Label III} setups.
Overall, we observe that classifying on a stronger fraud signal (\textit{Label II}) translates into better performance.
Also, these results seem to indicate that once a model is trained with a sufficiently strong signal, it is able to correctly label noisy data (AUC = 0.936) on \textit{Label III}.
This shows great promise in terms of the extensibility and applicability of our work.

\begin{table}
\centering
\caption{Average results for the neural network ensemble classifier comparing fraud scale labels. The standard deviation is shown in parenthesis.}
\label{table:ABC_ensemble_distribution}
\begin{tabular}{lccc}%{\columnwidth}{lXXX}
\toprule
\textbf{Scores} & \textbf{Accuracy} & \textbf{F1-Score}& \textbf{AUC}\\ \midrule

\textit{Label I}   & 0.8445(0.358) & 0.8438(0.036) & 0.9227(0.025)\\
\textit{Label II}  & 0.9014(0.034) & 0.9012(0.034) & 0.9602(0.022) \\
\textit{Label III} & 0.9077(0.052) & 0.8996(0.067) & 0.9358(0.051) \\

\bottomrule
\end{tabular}
\end{table}

%% file: sections/08_conclusion.tex
\section{Discussion and Future Work} 
\label{sec:conclusion}

%Evolutionary psychology has shown us that communities and individuals favor trust. In face-to-face communication, most people will perform only slightly better than random at detecting lies. 

In recent years, crowdfunding has emerged as a means of making personal appeals for financial support to members of the public.
These may be simple tasks such as a DIY project at home, or more complex ventures such as starting a new company or medical procedures.
The community trusts that the individual who requests support, whatever the task, is doing so without malicious intent.

However, time and again, fraudulent cases come to light, ranging from fake objectives to embezzlement.
Fraudsters often fly under the radar and defraud people of what adds up to tens of millions, under the guise of crowdfunding support, enabled by small individual donations.
Detecting and preventing fraud is thus an adversarial problem.
Inevitably, perpetrators adapt and attempt to bypass whatever system is deployed to prevent their malicious schemes.

In this work, we take the first step in studying the problem of fraudulent crowdfunding campaigns and detecting them at the time of publication.
We collect appropriate data from thousands of campaigns from different platforms and study fraud cases to better understand their characteristics.
Armed with this knowledge, we perform an annotation study to label hundreds of campaigns as fraud or not, with substantial overall annotation agreement.
We proceed to extract characteristics (features) from the text and image content included in each campaign, and compare these features with the associated label of the campaign.
%Finally, we use these features to train and test ensemble classifiers to automatically detect such campaigns in the wild.

The dataset we built is useful in training machine learning ensemble classifiers, which can take visual and textual cues from any crowdfunding campaign, and predict if the campaign is fraudulent or not when created, with satisfactory performance (up to AUC=0.96).
Indeed, there is room for improvement, especially regarding feature engineering and classifier complexity and tuning.
However, our results demonstrate that it is possible to detect fraudulent campaigns with high certainty, and allow crowdfunding platforms to remove them semi-automatically, i.e., can be marked for a more detailed inspection by an administrator.

In practice, we are proposing an automatic method that can help donors to have an indication of which of the campaigns they are viewing may be fraudulent.
With this method, we attempt to make the job of fraudsters harder, by proposing a better system than currently available.
In fact, in order to mitigate the risk of fraudsters catching up with the online model and what features it monitors for predicting fraud, we can explore different methods and timings of when to deliver the warning flag to a donor.

In terms of limitations while building this methodology, we attempted to reduce any bias that may have been introduced by the annotators.
During this process, we created checklists and standards into what would be defined as fraud, to minimize subjective bias.
A further unbiased way to conduct this study would be to rely exclusively on convicted cases of fraud, instead of relying on manual annotations of suspected cases.
However, this option would not provide enough examples to develop good enough machine learning models, and it would be again up to annotators to identify \textit{not-fraud} examples.
One solution that would further reduce the risk of bias is to increase the number of annotators that label each campaign.
But this is highly depended on resource availability.
%\bp{@Nicolas you mentioned yesterday something about more democratic ways to access as part of the topic of annotations, but I didn't understand what you meant.}
Finally, algorithmic bias could be reduced.
For example, poorly written campaigns by legitimate requestors who are uneducated or non-native English speakers can be mislabeled as fraud -- a clear source of bias.
Also, there may be limited examples of such campaigns, since these users may not be willing or comfortable to post a campaign in the first place. 
These aspects point to the problem of fair and balanced representation of characteristics in our training data and labels.

In the future, we plan to improve our classifier to take into account such sources of bias.
We also plan to test our classifier on unlabeled data of medically-related campaigns to investigate its capability detecting such fraud cases, which in the health domain can have a severe monetary and emotional impact on the defrauded.

%% file: ARCHIVE.bbl
%%% -*-BibTeX-*-
%%% Do NOT edit. File created by BibTeX with style
%%% ACM-Reference-Format-Journals [18-Jan-2012].

\begin{thebibliography}{42}

%%% ====================================================================
%%% NOTE TO THE USER: you can override these defaults by providing
%%% customized versions of any of these macros before the \bibliography
%%% command.  Each of them MUST provide its own final punctuation,
%%% except for \shownote{}, \showDOI{}, and \showURL{}.  The latter two
%%% do not use final punctuation, in order to avoid confusing it with
%%% the Web address.
%%%
%%% To suppress output of a particular field, define its macro to expand
%%% to an empty string, or better, \unskip, like this:
%%%
%%% \newcommand{\showDOI}[1]{\unskip}   % LaTeX syntax
%%%
%%% \def \showDOI #1{\unskip}           % plain TeX syntax
%%%
%%% ====================================================================

\ifx \showCODEN    \undefined \def \showCODEN     #1{\unskip}     \fi
\ifx \showDOI      \undefined \def \showDOI       #1{#1}\fi
\ifx \showISBNx    \undefined \def \showISBNx     #1{\unskip}     \fi
\ifx \showISBNxiii \undefined \def \showISBNxiii  #1{\unskip}     \fi
\ifx \showISSN     \undefined \def \showISSN      #1{\unskip}     \fi
\ifx \showLCCN     \undefined \def \showLCCN      #1{\unskip}     \fi
\ifx \shownote     \undefined \def \shownote      #1{#1}          \fi
\ifx \showarticletitle \undefined \def \showarticletitle #1{#1}   \fi
\ifx \showURL      \undefined \def \showURL       {\relax}        \fi
% The following commands are used for tagged output and should be
% invisible to TeX
\providecommand\bibfield[2]{#2}
\providecommand\bibinfo[2]{#2}
\providecommand\natexlab[1]{#1}
\providecommand\showeprint[2][]{arXiv:#2}

\bibitem[\protect\citeauthoryear{Abbasi, Albrecht, Vance, and Hansen}{Abbasi
  et~al\mbox{.}}{2012}]%
        {abbasi2012metafraud}
\bibfield{author}{\bibinfo{person}{Ahmed Abbasi}, \bibinfo{person}{Conan
  Albrecht}, \bibinfo{person}{Anthony Vance}, {and} \bibinfo{person}{James
  Hansen}.} \bibinfo{year}{2012}\natexlab{}.
\newblock \showarticletitle{Metafraud: A meta-learning framework for detecting
  financial fraud.}
\newblock \bibinfo{journal}{\emph{Mis Quarterly}} \bibinfo{volume}{36},
  \bibinfo{number}{4} (\bibinfo{year}{2012}).
\newblock


\bibitem[\protect\citeauthoryear{Belleflamme, Omrani, and Peitz}{Belleflamme
  et~al\mbox{.}}{2015}]%
        {belleflamme2015economics}
\bibfield{author}{\bibinfo{person}{Paul Belleflamme}, \bibinfo{person}{Nessrine
  Omrani}, {and} \bibinfo{person}{Martin Peitz}.}
  \bibinfo{year}{2015}\natexlab{}.
\newblock \showarticletitle{The economics of crowdfunding platforms}.
\newblock \bibinfo{journal}{\emph{Information Economics and Policy}}
  \bibinfo{volume}{33} (\bibinfo{year}{2015}), \bibinfo{pages}{11--28}.
\newblock


\bibitem[\protect\citeauthoryear{Bhattacharyya, Jha, Tharakunnel, and
  Westland}{Bhattacharyya et~al\mbox{.}}{2011}]%
        {bhattacharyya2011data}
\bibfield{author}{\bibinfo{person}{Siddhartha Bhattacharyya},
  \bibinfo{person}{Sanjeev Jha}, \bibinfo{person}{Kurian Tharakunnel}, {and}
  \bibinfo{person}{J~Christopher Westland}.} \bibinfo{year}{2011}\natexlab{}.
\newblock \showarticletitle{Data mining for credit card fraud: A comparative
  study}.
\newblock \bibinfo{journal}{\emph{Decision Support Systems}}
  \bibinfo{volume}{50}, \bibinfo{number}{3} (\bibinfo{year}{2011}),
  \bibinfo{pages}{602--613}.
\newblock


\bibitem[\protect\citeauthoryear{Biondi, Franzoni, and Poggioni}{Biondi
  et~al\mbox{.}}{2017}]%
        {IBM}
\bibfield{author}{\bibinfo{person}{Giulio Biondi}, \bibinfo{person}{Valentina
  Franzoni}, {and} \bibinfo{person}{Valentina Poggioni}.}
  \bibinfo{year}{2017}\natexlab{}.
\newblock \showarticletitle{{A Deep Learning Semantic Approach to Emotion
  Recognition Using the IBM Watson Bluemix Alchemy Language}}. In
  \bibinfo{booktitle}{\emph{ICCSA}}.
\newblock


\bibitem[\protect\citeauthoryear{Bond, Omar, Mahmoud, and Bonser}{Bond
  et~al\mbox{.}}{1990}]%
        {Bond1990}
\bibfield{author}{\bibinfo{person}{Charles~F. Bond}, \bibinfo{person}{Adnan
  Omar}, \bibinfo{person}{Adnan Mahmoud}, {and} \bibinfo{person}{Richard~Neal
  Bonser}.} \bibinfo{year}{1990}\natexlab{}.
\newblock \showarticletitle{Lie detection across cultures}.
\newblock \bibinfo{journal}{\emph{Journal of Nonverbal Behavior}}
  \bibinfo{volume}{14}, \bibinfo{number}{3} (\bibinfo{date}{Sep}
  \bibinfo{year}{1990}), \bibinfo{pages}{189--204}.
\newblock


\bibitem[\protect\citeauthoryear{Cecchini, Aytug, Koehler, and Pathak}{Cecchini
  et~al\mbox{.}}{2010}]%
        {cecchini2010detecting}
\bibfield{author}{\bibinfo{person}{Mark Cecchini}, \bibinfo{person}{Haldun
  Aytug}, \bibinfo{person}{Gary~J Koehler}, {and} \bibinfo{person}{Praveen
  Pathak}.} \bibinfo{year}{2010}\natexlab{}.
\newblock \showarticletitle{Detecting management fraud in public companies}.
\newblock \bibinfo{journal}{\emph{Management Science}} \bibinfo{volume}{56},
  \bibinfo{number}{7} (\bibinfo{year}{2010}), \bibinfo{pages}{1146--1160}.
\newblock


\bibitem[\protect\citeauthoryear{Chatfield, Simonyan, Vedaldi, and
  Zisserman}{Chatfield et~al\mbox{.}}{2014}]%
        {chatfield2014return}
\bibfield{author}{\bibinfo{person}{Ken Chatfield}, \bibinfo{person}{Karen
  Simonyan}, \bibinfo{person}{Andrea Vedaldi}, {and} \bibinfo{person}{Andrew
  Zisserman}.} \bibinfo{year}{2014}\natexlab{}.
\newblock \showarticletitle{Return of the devil in the details: Delving deep
  into convolutional nets}.
\newblock \bibinfo{journal}{\emph{arXiv preprint arXiv:1405.3531}}
  (\bibinfo{year}{2014}).
\newblock


\bibitem[\protect\citeauthoryear{Cohen}{Cohen}{1960}]%
        {cohen1960coefficient}
\bibfield{author}{\bibinfo{person}{Jacob Cohen}.}
  \bibinfo{year}{1960}\natexlab{}.
\newblock \showarticletitle{A coefficient of agreement for nominal scales}.
\newblock \bibinfo{journal}{\emph{Educational and psychological measurement}}
  \bibinfo{volume}{20}, \bibinfo{number}{1} (\bibinfo{year}{1960}),
  \bibinfo{pages}{37--46}.
\newblock


\bibitem[\protect\citeauthoryear{Cumming, Hornuf, Karami, and
  Schweizer}{Cumming et~al\mbox{.}}{2020}]%
        {cumming2016disentangling}
\bibfield{author}{\bibinfo{person}{Douglas~J Cumming}, \bibinfo{person}{Lars
  Hornuf}, \bibinfo{person}{Moein Karami}, {and} \bibinfo{person}{Denis
  Schweizer}.} \bibinfo{year}{2020}\natexlab{}.
\newblock \showarticletitle{Disentangling crowdfunding from fraudfunding}.
\newblock \bibinfo{journal}{\emph{Max Planck Institute for Innovation \&
  competition research paper}} \bibinfo{number}{16-09} (\bibinfo{year}{2020}).
\newblock


\bibitem[\protect\citeauthoryear{Dechow, Ge, Larson, and Sloan}{Dechow
  et~al\mbox{.}}{2011}]%
        {dechow2011predicting}
\bibfield{author}{\bibinfo{person}{Patricia~M Dechow}, \bibinfo{person}{Weili
  Ge}, \bibinfo{person}{Chad~R Larson}, {and} \bibinfo{person}{Richard~G
  Sloan}.} \bibinfo{year}{2011}\natexlab{}.
\newblock \showarticletitle{Predicting material accounting misstatements}.
\newblock \bibinfo{journal}{\emph{Contemporary accounting research}}
  \bibinfo{volume}{28}, \bibinfo{number}{1} (\bibinfo{year}{2011}),
  \bibinfo{pages}{17--82}.
\newblock


\bibitem[\protect\citeauthoryear{Duarte, Siegel, and Young}{Duarte
  et~al\mbox{.}}{2012}]%
        {duarte2012trust}
\bibfield{author}{\bibinfo{person}{Jefferson Duarte}, \bibinfo{person}{Stephan
  Siegel}, {and} \bibinfo{person}{Lance Young}.}
  \bibinfo{year}{2012}\natexlab{}.
\newblock \showarticletitle{Trust and credit: The role of appearance in
  peer-to-peer lending}.
\newblock \bibinfo{journal}{\emph{The Review of Financial Studies}}
  \bibinfo{volume}{25}, \bibinfo{number}{8} (\bibinfo{year}{2012}),
  \bibinfo{pages}{2455--2484}.
\newblock


\bibitem[\protect\citeauthoryear{{Gofundme Inc}}{{Gofundme Inc}}{2019}]%
        {gfm_pricing}
\bibfield{author}{\bibinfo{person}{{Gofundme Inc}}.}
  \bibinfo{year}{2019}\natexlab{}.
\newblock \bibinfo{title}{GoFundMe Pricing}.
\newblock \bibinfo{howpublished}{goFundme}.
\newblock
\newblock
\shownote{\url{http://gofundme.com/pricing}.}


\bibitem[\protect\citeauthoryear{{GoFundMe Inc}}{{GoFundMe Inc}}{2020}]%
        {gfmfraud}
\bibfield{author}{\bibinfo{person}{{GoFundMe Inc}}.}
  \bibinfo{year}{2020}\natexlab{}.
\newblock \bibinfo{title}{GoFundMe fraudulent campaigns}.
\newblock \bibinfo{howpublished}{GoFundMe}.
\newblock
\newblock
\shownote{\url{https://www.gofundme.com/c/safety/fraudulent-campaigns}.}


\bibitem[\protect\citeauthoryear{Gonzalez}{Gonzalez}{2014}]%
        {goFraudMe}
\bibfield{author}{\bibinfo{person}{Adrienne Gonzalez}.}
  \bibinfo{year}{2014}\natexlab{}.
\newblock \bibinfo{title}{GoFraudMe}.
\newblock \bibinfo{howpublished}{goFraudMe}.
\newblock
\newblock
\shownote{\url{http://gofraudme.com/}.}


\bibitem[\protect\citeauthoryear{He, Zhang, Ren, and Sun}{He
  et~al\mbox{.}}{2016}]%
        {he2016deep}
\bibfield{author}{\bibinfo{person}{Kaiming He}, \bibinfo{person}{Xiangyu
  Zhang}, \bibinfo{person}{Shaoqing Ren}, {and} \bibinfo{person}{Jian Sun}.}
  \bibinfo{year}{2016}\natexlab{}.
\newblock \showarticletitle{Deep residual learning for image recognition}. In
  \bibinfo{booktitle}{\emph{CVPR}}.
\newblock


\bibitem[\protect\citeauthoryear{Honnibal and Montani}{Honnibal and
  Montani}{2017}]%
        {honnibal2017spacy}
\bibfield{author}{\bibinfo{person}{Matthew Honnibal} {and}
  \bibinfo{person}{Ines Montani}.} \bibinfo{year}{2017}\natexlab{}.
\newblock \showarticletitle{spacy 2: Natural language understanding with bloom
  embeddings, convolutional neural networks and incremental parsing}.
\newblock \bibinfo{journal}{\emph{To appear}} \bibinfo{volume}{7},
  \bibinfo{number}{1} (\bibinfo{year}{2017}).
\newblock


\bibitem[\protect\citeauthoryear{Johnston}{Johnston}{2020}]%
        {fbi2020}
\bibfield{author}{\bibinfo{person}{Patrick Johnston}.}
  \bibinfo{year}{2020}\natexlab{}.
\newblock \bibinfo{title}{{FBI details new methods of fraud born amid the
  pandemic}}.
\newblock
  \bibinfo{howpublished}{\url{https://www.havredailynews.com/story/2020/04/22/local/fbi-details-new-methods-of-fraud-born-amid-the-pandemic/528576.html}}.
\newblock
\newblock
\shownote{Accessed: 2020-05-15.}


\bibitem[\protect\citeauthoryear{Joshi, Datta, Fedorovskaya, Luong, Wang, Li,
  and Luo}{Joshi et~al\mbox{.}}{2011}]%
        {joshi2011aesthetics}
\bibfield{author}{\bibinfo{person}{Dhiraj Joshi}, \bibinfo{person}{Ritendra
  Datta}, \bibinfo{person}{Elena Fedorovskaya}, \bibinfo{person}{Quang-Tuan
  Luong}, \bibinfo{person}{James~Z Wang}, \bibinfo{person}{Jia Li}, {and}
  \bibinfo{person}{Jiebo Luo}.} \bibinfo{year}{2011}\natexlab{}.
\newblock \showarticletitle{Aesthetics and emotions in images}.
\newblock \bibinfo{journal}{\emph{IEEE Signal Processing Magazine}}
  \bibinfo{volume}{28}, \bibinfo{number}{5} (\bibinfo{year}{2011}),
  \bibinfo{pages}{94--115}.
\newblock


\bibitem[\protect\citeauthoryear{King}{King}{2009}]%
        {king2009dlib}
\bibfield{author}{\bibinfo{person}{Davis~E King}.}
  \bibinfo{year}{2009}\natexlab{}.
\newblock \showarticletitle{Dlib-ml: A machine learning toolkit}.
\newblock \bibinfo{journal}{\emph{Journal of Machine Learning Research}}
  \bibinfo{volume}{10}, \bibinfo{number}{Jul} (\bibinfo{year}{2009}),
  \bibinfo{pages}{1755--1758}.
\newblock


\bibitem[\protect\citeauthoryear{Krizhevsky, Sutskever, and Hinton}{Krizhevsky
  et~al\mbox{.}}{2012}]%
        {krizhevsky2012imagenet}
\bibfield{author}{\bibinfo{person}{Alex Krizhevsky}, \bibinfo{person}{Ilya
  Sutskever}, {and} \bibinfo{person}{Geoffrey~E Hinton}.}
  \bibinfo{year}{2012}\natexlab{}.
\newblock \showarticletitle{Imagenet classification with deep convolutional
  neural networks}. In \bibinfo{booktitle}{\emph{NeurIPS}}.
\newblock


\bibitem[\protect\citeauthoryear{Landis and Koch}{Landis and Koch}{1977}]%
        {landis1977application}
\bibfield{author}{\bibinfo{person}{J~Richard Landis} {and}
  \bibinfo{person}{Gary~G Koch}.} \bibinfo{year}{1977}\natexlab{}.
\newblock \showarticletitle{An application of hierarchical kappa-type
  statistics in the assessment of majority agreement among multiple observers}.
\newblock \bibinfo{journal}{\emph{Biometrics}} (\bibinfo{year}{1977}),
  \bibinfo{pages}{363--374}.
\newblock


\bibitem[\protect\citeauthoryear{Lin, Prabhala, and Viswanathan}{Lin
  et~al\mbox{.}}{2013}]%
        {lin2013judging}
\bibfield{author}{\bibinfo{person}{Mingfeng Lin},
  \bibinfo{person}{Nagpurnanand~R Prabhala}, {and} \bibinfo{person}{Siva
  Viswanathan}.} \bibinfo{year}{2013}\natexlab{}.
\newblock \showarticletitle{Judging borrowers by the company they keep:
  Friendship networks and information asymmetry in online peer-to-peer
  lending}.
\newblock \bibinfo{journal}{\emph{Management Science}} \bibinfo{volume}{59},
  \bibinfo{number}{1} (\bibinfo{year}{2013}), \bibinfo{pages}{17--35}.
\newblock


\bibitem[\protect\citeauthoryear{Luca and Zervas}{Luca and Zervas}{2016}]%
        {luca2016fake}
\bibfield{author}{\bibinfo{person}{Michael Luca} {and}
  \bibinfo{person}{Georgios Zervas}.} \bibinfo{year}{2016}\natexlab{}.
\newblock \showarticletitle{Fake it till you make it: Reputation, competition,
  and Yelp review fraud}.
\newblock \bibinfo{journal}{\emph{Management Science}} \bibinfo{volume}{62},
  \bibinfo{number}{12} (\bibinfo{year}{2016}), \bibinfo{pages}{3412--3427}.
\newblock


\bibitem[\protect\citeauthoryear{McClanahan}{McClanahan}{2018}]%
        {forbesMedical}
\bibfield{author}{\bibinfo{person}{Carolyn McClanahan}.}
  \bibinfo{year}{2018}\natexlab{}.
\newblock \bibinfo{title}{People Are Raising USD650 Million On GoFundMe Each
  Year To Attack Rising Healthcare Costs}.
\newblock \bibinfo{howpublished}{Forbes}.
\newblock
\newblock
\shownote{\url{https://www.forbes.com/sites/carolynmcclanahan/2018/08/13/using-gofundme-to-attack-health-care-costs}.}


\bibitem[\protect\citeauthoryear{McHugh}{McHugh}{2012}]%
        {mchugh2012interrater}
\bibfield{author}{\bibinfo{person}{Mary~L McHugh}.}
  \bibinfo{year}{2012}\natexlab{}.
\newblock \showarticletitle{Interrater reliability: the kappa statistic}.
\newblock \bibinfo{journal}{\emph{Biochemia medica: Biochemia medica}}
  \bibinfo{volume}{22}, \bibinfo{number}{3} (\bibinfo{year}{2012}),
  \bibinfo{pages}{276--282}.
\newblock


\bibitem[\protect\citeauthoryear{Panigrahi, Kundu, Sural, and
  Majumdar}{Panigrahi et~al\mbox{.}}{2009}]%
        {panigrahi2009credit}
\bibfield{author}{\bibinfo{person}{Suvasini Panigrahi}, \bibinfo{person}{Amlan
  Kundu}, \bibinfo{person}{Shamik Sural}, {and} \bibinfo{person}{Arun~K
  Majumdar}.} \bibinfo{year}{2009}\natexlab{}.
\newblock \showarticletitle{Credit card fraud detection: A fusion approach
  using Dempster--Shafer theory and Bayesian learning}.
\newblock \bibinfo{journal}{\emph{Information Fusion}} \bibinfo{volume}{10},
  \bibinfo{number}{4} (\bibinfo{year}{2009}), \bibinfo{pages}{354--363}.
\newblock


\bibitem[\protect\citeauthoryear{Parsley}{Parsley}{2020}]%
        {ctm2020}
\bibfield{author}{\bibinfo{person}{David Parsley}.}
  \bibinfo{year}{2020}\natexlab{}.
\newblock \bibinfo{title}{{Captain Tom Moore: Just Giving blocks copycats over
  fears scammers are 'cashing in' on £28m NHS fundraising campaign}}.
\newblock
  \bibinfo{howpublished}{\url{https://inews.co.uk/inews-lifestyle/money/captain-tom-moore-war-hero-just-giving-copycats-scams-fundraising-nhs-2546244}}.
\newblock
\newblock
\shownote{Accessed: 2020-05-15.}


\bibitem[\protect\citeauthoryear{Pedregosa, Varoquaux, Gramfort, Michel,
  Thirion, Grisel, Blondel, Prettenhofer, Weiss, Dubourg,
  et~al\mbox{.}}{Pedregosa et~al\mbox{.}}{2011}]%
        {scikit}
\bibfield{author}{\bibinfo{person}{Fabian Pedregosa}, \bibinfo{person}{Ga{\"e}l
  Varoquaux}, \bibinfo{person}{Alexandre Gramfort}, \bibinfo{person}{Vincent
  Michel}, \bibinfo{person}{Bertrand Thirion}, \bibinfo{person}{Olivier
  Grisel}, \bibinfo{person}{Mathieu Blondel}, \bibinfo{person}{Peter
  Prettenhofer}, \bibinfo{person}{Ron Weiss}, \bibinfo{person}{Vincent
  Dubourg}, {et~al\mbox{.}}} \bibinfo{year}{2011}\natexlab{}.
\newblock \showarticletitle{{Scikit-learn: Machine learning in Python}}.
\newblock \bibinfo{journal}{\emph{Journal of Machine Learning Research}}
  \bibinfo{volume}{12}, \bibinfo{number}{Oct} (\bibinfo{year}{2011}),
  \bibinfo{pages}{2825--2830}.
\newblock


\bibitem[\protect\citeauthoryear{Pope and Sydnor}{Pope and Sydnor}{2011}]%
        {pope2011s}
\bibfield{author}{\bibinfo{person}{Devin~G Pope} {and}
  \bibinfo{person}{Justin~R Sydnor}.} \bibinfo{year}{2011}\natexlab{}.
\newblock \showarticletitle{What’s in a Picture? Evidence of Discrimination
  from Prosper. com}.
\newblock \bibinfo{journal}{\emph{Journal of Human resources}}
  \bibinfo{volume}{46}, \bibinfo{number}{1} (\bibinfo{year}{2011}),
  \bibinfo{pages}{53--92}.
\newblock


\bibitem[\protect\citeauthoryear{Popper and Lorenz}{Popper and Lorenz}{2020}]%
        {nyt2020}
\bibfield{author}{\bibinfo{person}{Nathaniel Popper} {and}
  \bibinfo{person}{Taylor Lorenz}.} \bibinfo{year}{2020}\natexlab{}.
\newblock \bibinfo{title}{{GoFundMe Confronts Coronavirus Demand}}.
\newblock
  \bibinfo{howpublished}{\url{https://www.nytimes.com/2020/03/26/style/gofundme-coronavirus.html}}.
\newblock
\newblock
\shownote{Accessed: 2020-05-15.}


\bibitem[\protect\citeauthoryear{S{\'a}nchez, Vila, Cerda, and
  Serrano}{S{\'a}nchez et~al\mbox{.}}{2009}]%
        {sanchez2009association}
\bibfield{author}{\bibinfo{person}{Daniel S{\'a}nchez}, \bibinfo{person}{MA
  Vila}, \bibinfo{person}{L Cerda}, {and} \bibinfo{person}{Jos{\'e}-Maria
  Serrano}.} \bibinfo{year}{2009}\natexlab{}.
\newblock \showarticletitle{Association rules applied to credit card fraud
  detection}.
\newblock \bibinfo{journal}{\emph{Expert systems with applications}}
  \bibinfo{volume}{36}, \bibinfo{number}{2} (\bibinfo{year}{2009}),
  \bibinfo{pages}{3630--3640}.
\newblock


\bibitem[\protect\citeauthoryear{Siering, Koch, and Deokar}{Siering
  et~al\mbox{.}}{2016}]%
        {siering2016detecting}
\bibfield{author}{\bibinfo{person}{Michael Siering},
  \bibinfo{person}{Jascha-Alexander Koch}, {and} \bibinfo{person}{Amit~V
  Deokar}.} \bibinfo{year}{2016}\natexlab{}.
\newblock \showarticletitle{Detecting fraudulent behavior on crowdfunding
  platforms: The role of linguistic and content-based cues in static and
  dynamic contexts}.
\newblock \bibinfo{journal}{\emph{Journal of Management Information Systems}}
  \bibinfo{volume}{33}, \bibinfo{number}{2} (\bibinfo{year}{2016}),
  \bibinfo{pages}{421--455}.
\newblock


\bibitem[\protect\citeauthoryear{Srivastava, Kundu, Sural, and
  Majumdar}{Srivastava et~al\mbox{.}}{2008}]%
        {srivastava2008credit}
\bibfield{author}{\bibinfo{person}{Abhinav Srivastava}, \bibinfo{person}{Amlan
  Kundu}, \bibinfo{person}{Shamik Sural}, {and} \bibinfo{person}{Arun
  Majumdar}.} \bibinfo{year}{2008}\natexlab{}.
\newblock \showarticletitle{Credit card fraud detection using hidden Markov
  model}.
\newblock \bibinfo{journal}{\emph{IEEE Transactions on dependable and secure
  computing}} \bibinfo{volume}{5}, \bibinfo{number}{1} (\bibinfo{year}{2008}),
  \bibinfo{pages}{37--48}.
\newblock


\bibitem[\protect\citeauthoryear{Szmigiera}{Szmigiera}{2018}]%
        {crowdfundingStatista}
\bibfield{author}{\bibinfo{person}{M Szmigiera}.}
  \bibinfo{year}{2018}\natexlab{}.
\newblock \bibinfo{title}{Crowdfunding Statistics and Facts}.
\newblock \bibinfo{howpublished}{Statista}.
\newblock
\newblock
\shownote{\url{https://www.statista.com/topics/1283/crowdfunding/}.}


\bibitem[\protect\citeauthoryear{{US Legal}}{{US Legal}}{2020}]%
        {fraudDef}
\bibfield{author}{\bibinfo{person}{{US Legal}}.}
  \bibinfo{year}{2020}\natexlab{}.
\newblock \bibinfo{title}{Fraud Law and Legal Definition}.
\newblock
\newblock
\newblock
\shownote{\url{https://definitions.uslegal.com/f/fraud}.}


\bibitem[\protect\citeauthoryear{Victor}{Victor}{2019}]%
        {nyt2019}
\bibfield{author}{\bibinfo{person}{Daniel Victor}.}
  \bibinfo{year}{2019}\natexlab{}.
\newblock \bibinfo{title}{{Woman and Homeless Man Plead Guilty in \$400,000
  GoFundMe Scam}}.
\newblock
  \bibinfo{howpublished}{\url{https://www.nytimes.com/2019/03/07/us/gofundme-homeless-scam-guilty.html}}.
\newblock
\newblock
\shownote{Accessed: 2020-05-15.}


\bibitem[\protect\citeauthoryear{Vonow}{Vonow}{2020}]%
        {theSun2020}
\bibfield{author}{\bibinfo{person}{Brittany Vonow}.}
  \bibinfo{year}{2020}\natexlab{}.
\newblock \bibinfo{title}{{LOWEST OF THE LOW Sick scammers are setting up
  GoFundMe accounts for fake coronavirus victims}}.
\newblock
  \bibinfo{howpublished}{\url{https://www.thesun.co.uk/news/11364340/scammers-fake-gofundme-coronavirus-victims/}}.
\newblock
\newblock
\shownote{Accessed: 2020-05-15.}


\bibitem[\protect\citeauthoryear{Wessel, Thies, and Benlian}{Wessel
  et~al\mbox{.}}{2016}]%
        {wessel2016emergence}
\bibfield{author}{\bibinfo{person}{Michael Wessel}, \bibinfo{person}{Ferdinand
  Thies}, {and} \bibinfo{person}{Alexander Benlian}.}
  \bibinfo{year}{2016}\natexlab{}.
\newblock \showarticletitle{The emergence and effects of fake social
  information: Evidence from crowdfunding}.
\newblock \bibinfo{journal}{\emph{Decision Support Systems}}
  \bibinfo{volume}{90} (\bibinfo{year}{2016}), \bibinfo{pages}{75 -- 85}.
\newblock


\bibitem[\protect\citeauthoryear{Xu, Lu, and Chau}{Xu et~al\mbox{.}}{2015}]%
        {xu2015p2p}
\bibfield{author}{\bibinfo{person}{Jennifer~J. Xu}, \bibinfo{person}{Yong Lu},
  {and} \bibinfo{person}{Michael Chau}.} \bibinfo{year}{2015}\natexlab{}.
\newblock \showarticletitle{P2P Lending Fraud Detection: A Big Data Approach}.
  In \bibinfo{booktitle}{\emph{ISI}}.
\newblock


\bibitem[\protect\citeauthoryear{You, Luo, Jin, and Yang}{You
  et~al\mbox{.}}{2016}]%
        {you2016building}
\bibfield{author}{\bibinfo{person}{Quanzeng You}, \bibinfo{person}{Jiebo Luo},
  \bibinfo{person}{Hailin Jin}, {and} \bibinfo{person}{Jianchao Yang}.}
  \bibinfo{year}{2016}\natexlab{}.
\newblock \showarticletitle{Building a large scale dataset for image emotion
  recognition: The fine print and the benchmark}. In
  \bibinfo{booktitle}{\emph{AAAI}}.
\newblock


\bibitem[\protect\citeauthoryear{Zeiler and Fergus}{Zeiler and Fergus}{2014}]%
        {zeiler2014visualizing}
\bibfield{author}{\bibinfo{person}{Matthew~D Zeiler} {and} \bibinfo{person}{Rob
  Fergus}.} \bibinfo{year}{2014}\natexlab{}.
\newblock \showarticletitle{Visualizing and understanding convolutional
  networks}. In \bibinfo{booktitle}{\emph{ECCV}}.
\newblock


\bibitem[\protect\citeauthoryear{Zhao, Gao, Jiang, Yao, Chua, and Sun}{Zhao
  et~al\mbox{.}}{2014}]%
        {zhao2014exploring}
\bibfield{author}{\bibinfo{person}{Sicheng Zhao}, \bibinfo{person}{Yue Gao},
  \bibinfo{person}{Xiaolei Jiang}, \bibinfo{person}{Hongxun Yao},
  \bibinfo{person}{Tat-Seng Chua}, {and} \bibinfo{person}{Xiaoshuai Sun}.}
  \bibinfo{year}{2014}\natexlab{}.
\newblock \showarticletitle{Exploring principles-of-art features for image
  emotion recognition}. In \bibinfo{booktitle}{\emph{MM}}.
\newblock


\end{thebibliography}
